\documentclass[aps, twocolumn,floats, showpacs]{revtex4}
\usepackage{amsmath,amssymb,graphicx}
\usepackage{epsfig}
\usepackage{graphicx}
\usepackage{enumerate}

\newcommand{\beq}{\begin{equation}}
\newcommand{\eeq}{\end{equation}}

\newcommand{\bea}{\begin{eqnarray}}
\newcommand{\eea}{\end{eqnarray}}

\begin{document}
\title{Nonlinear thermal control in an $N$-terminal junction}
\author{Dvira Segal}
\affiliation{Chemical Physics Theory Group, Department of Chemistry, University of Toronto,
80 Saint George St. Toronto, Ontario, Canada M5S 3H6}

\date{\today}
\begin{abstract}
We demonstrate control over heat flow in  an $N$-terminal molecular
junction. Using simple model Hamiltonians  we show that the heat current
through two terminals can be tuned, switched, and amplified, by the
temperature and coupling parameters of external gating reservoirs.
We discuss two models: A fully harmonic system, and a model
incorporating anharmonic interactions. For both models  the control
reservoirs induce thermal fluctuations of the transition elements
between molecular vibrational states. We find that a fully harmonic
model does not show any controllability, while for an anharmonic
system the conduction properties of the junction strongly depend on
the parameters of the gates. Realizations of the model system within
nanodevices and macromolecules are discussed.
\end{abstract}


\maketitle

Control over vibrational energy flow in nanoscale structures and
single molecules is a long standing goal in many parts of physical
science and nanotechnology. Historically, intramolecular vibrational
redistribution (IVR) was a topic of great interest in chemistry and
physics. IVR processes must be reckoned for understanding, and
ultimately controlling, molecular dynamics and chemical kinetics \cite{Uzer}.
The efficiency of these processes is the basic assumption behind the
well validated 
RRKM reaction rate theory \cite{RRKM, Wolynes}. From a different perspective, the unexpected
results of the computer experiment of Fermi-Pasta-Ulam \cite{FPU},
showing no equipartition of energy among normal modes in harmonic
chains including small nonlinear terms, lead to extensive research
of IVR in nonlinear systems \cite{FPURev}.

Recurrent theoretical interest in this field is due to the
impressive progress in probing thermal properties of
nanoscale systems such as nanotubes \cite{Kim,Fujii,Yu}, self
assembled monolayers \cite{Braun,Segalman,Dlott}, and thin films \cite{films},
and due to the development of more tunable systems  \cite{RectifE}.
Recent progress in molecular electronics and nanomechanics has
raised further interest in exploring mechanisms of energy flow
in nano-level systems. In molecular
electronics, local heating of nanoscale devices might cause
structural instabilities undermining the junction integrity
\cite{HeatSegal, HeatDiVentra, Galperin}. Engineering good thermal
contacts and cooling of the the junction are necessary for a stable
operation mode. Minimization of mechanical devices,
e.g. refrigerators \cite{Saira} and pumps \cite{Pump}, to the
molecular scale is a topic of great interest for technologies such
as chemical sensing, power generation and energy conversion
\cite{films,Cahill,Blencowe,Majumdar}. In this context, it is crucial to
understand, and ultimately control, the dynamics of phonons in
nanoscale structures, or analogously, vibrational modes in molecular
systems.

\begin{figure}[htbp]
\vspace{-10mm}
\hspace{-15mm}
{\hbox{\epsfxsize=110mm \epsffile{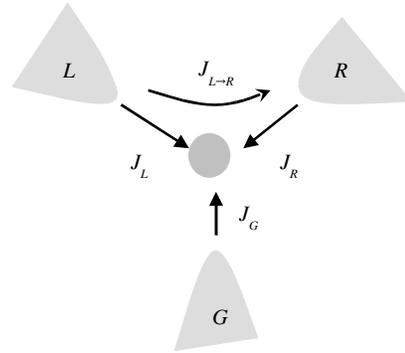}}}
\vspace{-85mm}
\caption{A schematic representation of the model studied in this work,
exemplified on a three-bath system.
The molecular units (central part) interacts with multiple thermal reservoirs
maintained at constant temperatures.
Heat current is defined as positive when flowing into the central molecular segment.
}
\label{Fig0}
\end{figure}

The heat conduction properties of molecular chains can be tuned
either by manipulating the internal molecular parameters, or by
externally gating the system. We refer to the first mode of control
as "static", or "internal", while, as we explain below, we consider gating as a
"dynamical", or an "external" control scheme.

Static control 
is realized
by adjusting internal system parameters, e.g. atomic masses and interatomic
potential energy, with the goal of increase/decrease of the system
thermal conductivity \cite{CasatiChaos}. This type of control
problem has been extensively discussed in the
context of Fourier transport. Here the main challenge is to identify
the necessary and sufficient conditions for the validity of the
Fourier law of heat conductivity, $J\propto -\nabla T$, in low
dimensions and for quantum systems
\cite{Bonetto,Lepri,Gemmer,Pereira}. By Engineering the molecular
system, one can also build functional devices, for example, a thermal
rectifier, where the nanojunction allows heat flux in one direction,
while it acts like an insulator when the temperature
gradient is reversed \cite{Casati1, Casati2,Rectif,RectifE}.

In this context we would like to emphasize that tuning the
thermal conductance of {\it harmonic} chains is also feasible,
though these systems demonstrate abnormal (non-Fourier) thermal
current. In the language of the thermal Landauer formula
\cite{Rego}, this can be accomplished by manipulating the transmission
coefficient for phononic heat flow through the device,
e.g. by introducing impurities into the structure \cite{heatcond}.

In this paper we present a simple model that illustrates
an  {\it external} control over thermal current in molecular systems.
The generic setup includes a molecule (subsystem) coupled to $N$
thermal baths of different temperatures.
Heat current flowing through the system may be
modified by a control reservoir, where in general, more than one
gate may couple to the subsystem. Typical control parameters are the
gate temperature and its coupling to the molecular unit. A schematic
representation of a three-bath scenario is shown in Fig. \ref{Fig0}.

We consider two realizations of this standard setup: (i) Artificial
nanodevices operating as thermal amplifiers or transistors
\cite{Casatixxx}. Here the generic system includes three segments,
source ($L$) drain ($R$) and gate ($G$), following the notation
 used in semiconductor transistors, where heat flow between the
source and  drain is controlled by the gate terminal. 
This device may be realized by fabricating branched nanotubes with T, Y,
and X shapes  \cite{Tshape}.
(ii) Macromolecules, e.g. proteins and dendriemds with spatially
separated sidegroups. Energy may be funneled between the molecular
groups by a control unit ($G$), for example a protein, that can
temporarily bind to the system. We refer to the energetically hot
group as $L$, while the $R$ group is the energy accepting sidegroup.

The role of the gate ($G$) may be modeled as inducing thermal
fluctuations of the $L$ - $R$ transition element \cite{Trinkunas},
thus we refer to this mode of control as dynamical.
The control element can also be identified as the solvent itself, modulating
system parameters. Inserting a molecule in different mediums
may therefore modify its heat conduction properties. In this context this work
provides a simple framework for investigating IVR in solutions
\cite{Schwarzer2}.

The main question to be addressed in this paper is what is the role
of anharmonic interactions, specifically,
{\it nonbilinear} molecule-surface interactions,
in controlling the thermal properties of a gated system.
It is widely accepted that nonlinear interactions are
essential for showing normal (Fourier) transport in molecular chains
\cite{Lepri,Dhar}. Anharmonic interactions are also necessary for
bringing in rectifying behavior \cite{Casati1, Casati2, Rectif}, and
for manifesting nonlinear thermal conductance characteristics  \cite{NDR,Casatixxx}. We
show next that anharmonic interactions are also a crucial element
for realizing dynamical control of heat current.

%
We consider two models. The first  system is a
prototype for transport in harmonic chains. The second model
incorporates anharmonic interactions in the molecule, and also
assumes nonbilinear system-bath couplings. For both models we focus
on two quantities: (i) We calculate the heat current at the terminal
$\nu$, $J_{\nu}$, and (ii) we investigate the net heat current
flowing between two surfaces, $J_{\nu \rightarrow \mu}$.
The objective of our calculation is to demonstrate that these
quantities strongly depend on the parameters of the gate reservoirs
(temperature and energetics) for the anharmonic model only.

We begin with the harmonic model. In this case
both the reservoirs (inverse temperatures $\beta_{\nu}=T_{\nu}^{-1}$, $\nu=1...N$) 
and the molecular unit are modeled
by a set of non interacting bosons coupled via a bilinear term.
For simplicity, we assume that heat transfer in the molecular unit is dominated by a 
specific single mode,

\bea
H_{bb}&=&\omega_0 b^{\dagger}b +
\sum_{j,\nu}\omega_{j}a_{j,\nu}^{\dagger} a_{j,\nu}
\nonumber\\&+&(b^{\dagger}+b) \sum_{j,\nu} \lambda_{j,\nu}(a_{j,\nu}^{\dagger}+a_{j,\nu}).
\label{eq:Hbb}
\eea
Here $b^{\dagger}$ ($b$) are creation (annihilation) operators for the molecular mode
of frequency $\omega_0$.
Similarly, $a_{j,\nu}^{\dagger}$ ($a_{j,\nu}$) are the operators for the
mode $j$ of the $\nu$ reservoir. Since system-bath interaction is bilinear,
the model Hamiltonian can be exactly diagonalized,
to be represented in terms of a set of noninteracting phonons.
We refer to model (\ref{eq:Hbb}) as the boson-boson (b-b) Hamiltonian.

The dynamics of the subsystem can be
exactly solved using various techniques, e.g. the generalized
Langevin equation \cite{heatcond,Dhar} and Master equation formalism
\cite{Rectif,NDR,Lin}. The result of these calculations is the
"thermal Landauer formula" \cite{heatcond,Rego}, where in the
classical limit the heat current (for $N=2$) linearly depends
on the temperature difference between the two thermal baths.
We briefly follow here the derivation within the Master equation
formalism, generalizing the results of Refs. \cite{Rectif,NDR} for an $N$
terminal system.

Under the assumption of weak system-bath interactions, going into
the Markovian limit, the probabilities $P_n$ to occupy the $n$ state
of the molecular oscillator satisfy the Master equation
\cite{Lin,NDR},
\bea
\dot P_n &=& n k _u  P_{n-1}(t) \nonumber\\
&+&(n+1)k_dP_{n+1}(t)-[nk_d+(n+1)k_u]P_n(t).
\label{eq:Pn}
\eea
The nonadiabatic relaxation and excitation rates, $k_d$
and $k_u$ respectively, are given by summing up contributions from
each reservoir, as no correlations exist between the different baths
($\nu=1..N$),
\bea
k_u=\sum_{\nu } {k_u^{\nu}}; \,\,\ k_d=\sum_{\nu}{k_d^{\nu}}.
\label{eq:rate1}
\eea
It can be shown that
\bea
k_d^{\nu}= \Gamma_{\nu}(\omega_0)(1+\bar n_{\nu}(\omega_0))
;\,\,\, k_u^{\nu}= \Gamma_{\nu}(\omega_0)\bar n_{\nu}(\omega_0),
\label{eq:rate2}
\eea
where
\bea \Gamma_{\nu}(\omega_0)=2\pi \sum_{j\in \nu} \lambda_j^2
\delta(\omega_j-\omega_0). \eea
Here $\bar n_{\nu}(\omega) =[e^{\omega/T_{\nu}}-1]^{-1}$ is the
Bose-Einstein distribution function for the $\nu$ reservoir.
The heat current properties of this model are obtained from the
steady state solution of Eq. (\ref{eq:Pn}) with the nonadiabatic rates
(\ref{eq:rate1})-(\ref{eq:rate2}). The steady state heat flux at the
$\nu$ terminal is given by calculating  the difference between
 heat flow from the $\nu$ bath into the the molecular mode,
 leading to vibrational excitations within the molecule,
and the outgoing molecule-reservoir energy current, resulting from relaxation
processes inside the molecule,
\bea
J_{\nu}^{(bb)}=-\omega_0 \sum_{n} n
(k_d^{\nu}P_n-k_u^{\nu}P_{n-1} ).
\eea
The current is defined  positive when flowing from the
contact into the molecule. In the classical limit ($T_{\nu}> \omega_0$) we get
\cite{NDR}
\bea
J_{\nu}^{(bb)}&=& \frac{\Gamma_{\nu}  \sum_{\mu} \Gamma_{\mu}
(T_{\nu}-T_{\mu}) }{\sum_{\mu} \Gamma_{\mu}}.
\label{eq:Jnubb}
\eea
Considering this expression, we can identify the directed current
$\nu \rightarrow \mu$ as
\bea
J_{\nu \rightarrow \mu}^{(bb)}&=& \frac{\Gamma_{\nu}
\Gamma_{\mu} (T_{\nu}-T_{\mu}) } {\sum_{\mu} \Gamma_{\mu}}.
\label{eq:Jnumubb}
\eea
We refer next to two specific terminals as source ($L$) and drain
($R$), while all other $N-2$ baths are referred to as gates ($G$).
When  currents in the $G$ terminals are
zero, $J_{\nu\neq L,R}=0$,  the gates acquire the same temperature in steady state,
$T_G=(\Gamma_LT_L+\Gamma_RT_R)/(\Gamma_L+\Gamma_R)$. The current at
the $L/R$ contact is then given by
\bea
J_{L}^{(bb)}=\frac{\Gamma_L \Gamma_R}{\Gamma_L+\Gamma_R}(T_L-T_R) ; \,\,\,\ J_R=-J_L,
\label{eq:JLbb}
\eea
which is the same result as obtained when $\Gamma_G=0$, see Eq. (\ref{eq:Jnubb}).
We also find that the current $J_{L\rightarrow
R}$ decays with the number of thermal reservoirs as
\bea
J_{L\rightarrow R}^{(bb)} = \frac{\Gamma_L \Gamma_R} {\sum_{\nu}\Gamma_{\nu}} (T_L-T_R); \,\,\,\,\,\ (J_G=0),
\label{eq:JLRbb}
\eea
due to additional decay channels. To summarize, we find that in the
harmonic limit the effect of gate terminals is simply to
effectively increase the broadening $\Gamma$, while the
gates' temperatures can not modify the current in the system.
Thus, there is no control over the heat
dissipated (or absorbed) from the contacts in a purely harmonic
system.

Next we show that in a model consisting nonlinear interactions heat
current can be strongly controlled by the temperature of a gate terminal.
As a case study we consider the spin-boson (s-b) model, generalized
to include $N$ bosonic reservoirs  (creation operators
$a_{j,\nu}^{\dagger}$, $\nu=1...N$) linearly coupled to a spin (two-level)
system,
\bea
H_{sb}&=& \frac{\Delta}{2} \sigma_x +\frac{\omega_0}{2} \sigma_z+
\sum_{j,\nu}\omega_ja_{j,\nu}^{\dagger} a_{j,\nu}
\nonumber\\
&+& \frac{\sigma_z}{2}\sum_{j,\nu} \kappa_{j,\nu}(a_{j,\nu}^{\dagger}+a_{j,\nu}).
\label{eq:Hsb}
\eea
Here $\omega_0$ is the energy difference between the spin levels
with tunneling splitting $\Delta$. In this model internal molecular
anharmonicity is introduced  by truncating the spectrum of the single
molecular mode to include only the lowest two energy states. We do
not allow for other phonon-phonon scattering processes, e.g. umklapp
processes, that can lead to normal conductivity as in the Peierls
model \cite{Peierls}. Using the small polaron transformation
\cite{Mahan} it can be shown that this model represents a molecular
mode coupled {\it nonbilinearly} to the harmonic reservoirs
\cite{Rectif},
\bea H_{sb}&= &\frac{\omega_0}{2} \sigma_z + \sum_{j,\nu} \omega_j
a_{j,\nu}^{\dagger}a_{j,\nu} \nonumber\\ &+&\frac{\Delta}{2}(
e^{i\Omega} \sigma_+ + e^{-i\Omega} \sigma_-) + H_{shift}.
\label{eq:Hpolaron} \eea
Here $\Omega=\sum_{\nu} \Omega_{\nu}$;
$\Omega_{\nu}=i\sum_{j}\frac{\kappa_{j,\nu}}{\omega_j}(a_{j,\nu}^{\dagger}-a_{j,\nu}
)$, $H_{shift}=\sum_{j,\nu}\frac{-\kappa_{j,\nu}^2}{4\omega_j}$ is an energy
shift henceforth incorporated into the zero order energies. Eq.
(\ref{eq:Hpolaron}) shows that the role of the thermal reservoirs is
to modulate the transition elements between molecular vibrational
levels. 
The important feature of this model is that system-bath
couplings (Eq. (\ref{eq:Hpolaron})) are multiplicative in the bath degrees of freedom, 
rather than additive as is the linear harmonic model (\ref{eq:rate1}). 
We do not distinguish in this model between the role of the different reservoirs 
(source, drain and gates). One could construct variants of this model, 
where the gates interact in a distinct functional form.
For small $\Delta$ the Hamiltonian leads again to nonadiabatic dynamics, Eq.
(\ref{eq:Pn}), with $n=0,1$. The and rate constants are given by
\bea
k_u=\frac{\Delta^2}{4}C(-\omega_0);  \,\,\, k_d=\frac{\Delta^2}{4}C(\omega_0).
\eea
Here
\bea
C(\omega_0)&=&\int_{-\infty}^{\infty} dt e^{i\omega_0t} C(t),
\nonumber\\
C(t)&=&\Pi_{\nu} C_{\nu}(t); \,\,\,\,
C_{\nu}(t)=\langle  e^{i\Omega_{\nu}(t)}
e^{-i\Omega_{\nu}(0)}\rangle _{\nu}.
\eea
The trace is performed over the $\nu$ reservoir degrees of freedom.
For convenience, in what follows we disregard the prefactor $(\Delta/2)^2$.
Using the convolution theorem, it can be shown that the function $C(\omega)$
can be decomposed in terms of the $N$ reservoirs correlation functions,
\bea
&&C(\omega_0)=\int_{-\infty}^{\infty} d\omega_1
\int_{-\infty}^{\infty} d\omega_2...
\int_{-\infty}^{\infty} d\omega_{N-1}
\nonumber\\
&&\times C_{1}(\omega_1)C_{2}(\omega_2)...
C_N(\omega_0-\omega_1-\omega_2-...-\omega_{N-1})
\nonumber\\
\eea
where $C_{\nu}(\omega)=\int_{-\infty}^{\infty}e^{i\omega
t}C_{\nu}(t)dt$ is identified as the rate constant affected from the $\nu$ thermal bath.
%
The heat flux at the $\nu$ terminal can  be formally written for an
$N$ terminal system by considering combinations of all processes in
which the reservoirs exchange energy with the subsystem \cite{Rectif},
\bea
&&J_{\nu}^{(sb)}=- \int_{-\infty}^{\infty} \omega_{\nu}
d\omega_1 d \omega_2... d\omega_{\nu}... d \omega_{N-1}
\Pi_{k\neq \nu, N}
C_{k}(\omega_k) \times
\nonumber\\
&&\big[ C_{ \nu}(\omega_{\nu})
C_{N}(\omega_0-\omega_1-\omega_2...-\omega_\nu...-\omega_{N-1})P_1 -
\nonumber\\
&& C_{ \nu}(-\omega_{\nu}) C_N(- \omega_1-\omega_2...
+\omega_{\nu}...-\omega_{N-1}-\omega_0) P_0\big].
\nonumber\\
\label{eq:Jsb}
\eea
For clarity, we include the explicit expression for the heat flux in
a three-terminal junctions, measured at terminal '1',
\bea 
&&J_1^{(sb)}(N=3)=
\nonumber\\
&&- \int_{-\infty}^{\infty} \omega_1  d\omega_1 \int_{-\infty}^{\infty}
d\omega_2 
\large[C_1(\omega_1)C_2(\omega_2)C_3(\omega_0-\omega_1-\omega_2)P_1
\nonumber\\
&&- C_1(-\omega_1)C_2(\omega_2)C_3(-\omega_0+\omega_1-\omega_2)P_0\large].
 \eea
The  population of the spin levels is given by
$P_0=1-P_1=C(\omega_0)/[C(\omega_0)+C(-\omega_0)]$. Assuming strong
coupling, going into the high temperature classical limit, ($T_{\nu}
> \omega_0$), the kernel $C(t)$ can be calculated in the short time
limit \cite{Mahan}, 
\bea
C_{\nu}(t)=e^{-\phi_{\nu}(t)}; \,\,\, \phi_{\nu}(t)=iE_{M}^{\nu}t +T_{\nu}E_M^{\nu}t^2,
\eea
where $E_M^{\nu}= \sum_{j}\frac{\kappa_{j,\nu}^2}{\omega_j}$ is the
reorganization energy of the $\nu$ reservoir. In frequency domain we
find that
\bea
&&C(\omega_0)=\sqrt{\frac{\pi}{T_M E_M } }
\exp\left[-\frac{(\omega_0-E_M)^2}{4T_ME_M}\right],
\nonumber\\
&&C_{\nu}(\omega)=\frac{1}{\sqrt{2T_{\nu}E_M^{\nu}}}\exp\left[-\frac{(\omega-E_M^{\nu})^2}{4 T_{\nu}E_M^{\nu}}\right],
\label{eq:C}
\eea
where $E_M$ is the total ($N$ baths) reorganization energy and $T_M$ is an
effective temperature for the subsystem,
\bea
E_{M}=\sum_{\nu}E_M^{\nu};
\nonumber\\
T_M=\frac{\sum_{\nu} E_M^{\nu}T_{\nu}}{E_M}. \eea
Integrating Eq. (\ref{eq:Jsb}) utilizing (\ref{eq:C}) yields the heat
current at the $\nu$ contact (ignoring a multiplicative numeric
factor of $\sqrt{4 \pi}$),
\bea
J_{\nu}^{(sb)}= E_{M}^{\nu}\frac{ \sum_{\mu
}E_{M}^{\mu}(T_{\nu}-T_{\mu}) }{(E_MT_M)^{3/2}(1+e^{\omega_0/T_M})}
e^{-\frac{(\omega_0-E_M)^2}{4E_MT_M}}.
\label{eq:Jnusb}
\eea
The current flowing through the system between the terminals $\nu$
and $\mu$ is given by
\bea
J_{\nu\rightarrow \mu}^{(sb)}= \frac{E_M^{\nu}
E_{M}^{{\mu}}(T_{\nu}-T_{\mu})} {(E_MT_M)^{3/2}(1+e^{\omega_0/T_M})}
e^{-\frac{(\omega_0-E_M)^2}{4E_MT_M}}.
\label{eq:Jnumusb}
\eea
The temperatures of the gating terminals and their couplings to the
molecule appear in a nontrivial way in this expression, leading to
strong controllability, as opposed to the harmonic results, Eqs.
(\ref{eq:Jnubb}) and (\ref{eq:Jnumubb}).

We exemplify control over the heat current in the system by
studying a source-drain-gate situation, where two reservoirs are
considered as source ($L$) and drain ($R$), while $N-2$ baths are
identified as gates ($G$). Under the condition of zero current in
the gating terminals ($J_{G}=0$), their temperatures can be
determined self consistently to yield
$T_G=T_M=(E_M^LT_L+E_M^RT_R)/(E_M^L+E_M^R)$.
The current at the source/drain contact then becomes
\bea
J_{L}^{(sb)}&=&-J_R^{(sb)}
\nonumber\\
&=& (T_L-T_R)\frac{E_M^L E_M^R}{E_M^L+E_M^R}\frac{E_M}{(E_MT_G)^{3/2}}
\frac{ e^{-\frac{(\omega_0-E_M)^2}{4E_MT_G}} }
{(1+e^{\omega_0/T_G})}.
\nonumber\\
\label{eq:JLsb}
\eea
When all gates evenly couple to the subsystem, $E_M^{\nu}=E_M^0$, and
for $E_M=N E_M^0 > \omega_0$, we find that the current exponentially decays with
$N$ (corrected by a power law), $J_L^{(sb)} \propto \Delta T
N^{-1/2}e^{-NE_M^0/4T_G}$, $\Delta T=T_L-T_R$.  We can also
calculate the $L\rightarrow R$ current, again taking $J_G=0$,
\bea
J_{L\rightarrow R}^{(sb)}=
(T_L-T_R)\frac{E_M^LE_M^R}{(E_MT_G)^{3/2}}
\frac{ e^{-\frac{(\omega_0-E_M)^2}{4E_MT_G}} }
{(1+e^{\omega_0/T_G})}.
\label{eq:JLRsb}
\eea
In the limit of  strong coupling, $N E_M^0 > \omega_0$, we find
that $J_{L\rightarrow R}^{(sb)} \propto \Delta T
N^{-3/2}e^{-NE_M^0/4T_G}$. Therefore, the temperature $T_G$
serves as an effective activation temperature, exponentially
enhancing the directed current, while $E_M$ is the energy gap for
transport. Note that $J_{L}/J_{L\rightarrow R}=E_M/(E_M^{L}+E_M^{R})$, 
in analogy with the behavior of the
fully harmonic model, Eq. (\ref{eq:JLbb})-(\ref{eq:JLRbb}). 


%
\begin{figure}[htbp]
\vspace{0mm} \hspace{0mm} {\hbox{\epsfxsize=85mm
\epsffile{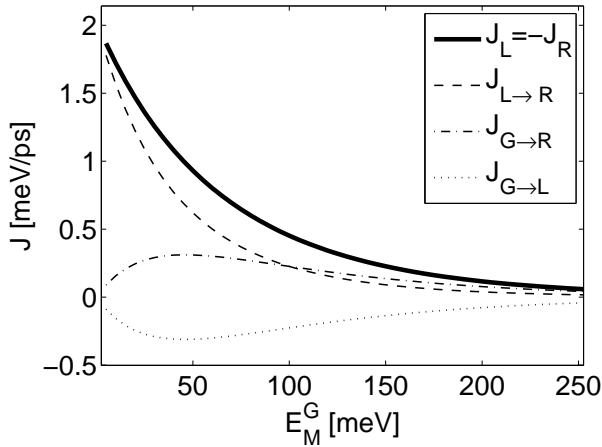}}} \caption{Control over heat current in a
three-terminal configuration under the condition of zero current at
the gate $J_G$=0. The heat current at the $L$ contact (full)
strongly decays with $E_M^G$. Also shown are the directed currents
$J_{L\rightarrow R}$ (dashed), $J_{G\rightarrow R}$ (dashed-dotted),
and $J_{G\rightarrow L}$ (dotted). Other parameters are $T_L$=300K,
$T_R$=200K, $T_G$=250K, $E_M^L=E_M^R=50$ meV.} \label{Fig1}
\end{figure}
\begin{figure}[htbp]
\vspace{0mm} \hspace{0mm} {\hbox{\epsfxsize=85mm
\epsffile{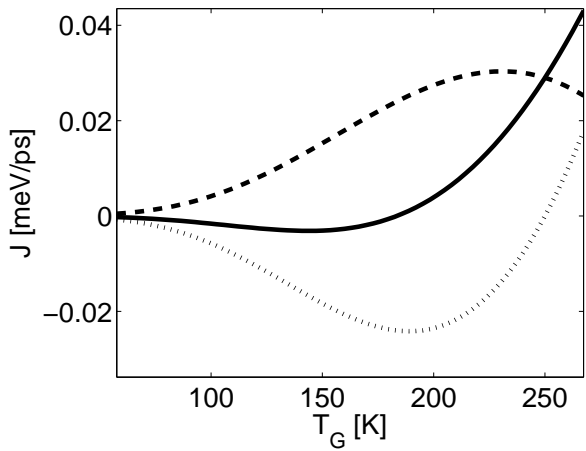}}} \caption{Switching the heat current in a
three-terminal setup: The system behaves as an insulator for low
control temperatures, while for high $T_G$ the system becomes a good
thermal conductor. $J_L$ (dashed) and $-J_R$ (full) are equal at
$T_G=250 K$, where $J_G$ (dotted) diminishes. 
Other parameters are $T_L$=300K, $T_R$=200K,
$E_M^L=E_M^R=50$ meV, $E_M^G=300$ meV.}
 \label{Fig2}
\end{figure}
%
%
\begin{figure}[htbp]
\vspace{0mm} \hspace{0mm} {\hbox{\epsfxsize=85mm
\epsffile{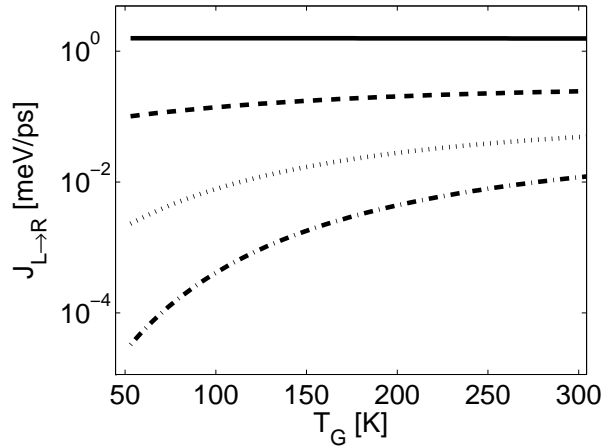}}} \caption{Amplification of the heat
current in a three-terminal configuration. The current
$J_{L\rightarrow R}$ can be strongly manipulated by the control bath
at strong coupling. $T_L$=300K, $T_R$=200K, $E_M^L=E_M^R=50$ meV,
$E_M^G$=10 meV (full), 100 (dashed), 200 (dotted) and 300 meV
(dashed-dotted).} \label{Fig3}
\end{figure}
%

We present next  numerical results calculated by applying Eqs.
(\ref{eq:Jnusb})-(\ref{eq:Jnumusb}) on a three-terminal ($L$, $R$
and $G$) configuration. The control parameters are the
coupling strength of the gate to the subsystem, given in terms of
reorganization energy $E_M^G$ and the temperature of the $G$
reservoir. For simplicity we take $\Delta/2=1$ meV.

Figure \ref{Fig1} displays results in the absence of
(net) energy flow between the gate and the subsystem. Taking all coupling strengths
to be equal sets the gate temperature to $T_G=(T_L+T_R)/2$. 
Simulating Eq. (\ref{eq:JLsb}) we find that the current at the $L/R$
interfaces strongly decays with $E_M^G$. We also show strong control over
the directed $L\rightarrow R$ current using Eq. (\ref{eq:JLRsb}).

Next we modulate the temperature of the gate reservoir, and manifest
that the system can act like a good thermal conductor, or an
insulator, depending on the gate temperature. Figure
\ref{Fig2} shows that the system behaves as a bad thermal conductor
at low gating temperature $T_G\sim 100 K$, while at $T_G=250 K$
(where $J_G=0$) the current is amplified by a factor of $\sim$10.

Finally, in Figure \ref{Fig3} we analyze the directed heat flux
$J_{L\rightarrow R}$ as a function of the gate temperature. We show
strong amplification of current, of three orders of magnitudes, 
when the gate terminal is strongly coupled to
the molecular mode (dashed-dotted line). In contrast, at weak
coupling (full line), $J_{L \rightarrow R}$ is insensitive to the
presence of the $G$ channel.
As discussed above,  the model Hamiltonian
(\ref{eq:Hsb}) could be also introduced
for describing transport between two sidegroups in a
macromolecule immersed in a solvent. In this context, Fig.
\ref{Fig3} reveals that the solvent may enhance the $L$ to $R$
current, in analogy with the effect of solvent assisted IVR
\cite{Schwarzer2}.


In conclusion, we have presented here two models for the study of thermal transport
in gated systems: a purely harmonic system, and an anharmonic model.
In both cases the gate terminals induce thermal fluctuations of the
transition elements between the molecular vibrational states,
leading to dynamical control of heat current.
We (trivially) found that the purely harmonic setup cannot bring in a
gated behavior. In contrast, in the spin-boson model, incorporating anharmonic
interactions, the system can behave either as an insulator, or as a good thermal
conductor, depending on the gates parameters: temperatures and molecule-bath
coupling strength (binding energy using proteins terminology).

We may also explore other variants of the anharmonic model. For example,
the central molecular unit can bilinearly-weakly couple to the $L$ and $R$
surfaces, while the gates may couple nonbilinearly-strongly to the subsystem. 
Such models should basically display the same characteristics as discussed above.

The effects described in this paper may be also studied using
classical molecular dynamics (MD) simulations \cite{Casatixxx}. The
advantage of our formalism over such Langevin equation treatment
is twofold: (i) Nonbilinear system-surface couplings are difficult to
simulate within Langevin dynamics, leading to a coordinate
dependent friction coefficient \cite{Carmeli}. 
(ii) The net heat current between two reservoirs, $J_{\nu
\rightarrow \mu}$, cannot be resolved within MD simulations, 
only the current at each terminal can be directly obtained. In
contrast, one can clearly identify these currents in the
analytical expression (\ref{eq:Jnumusb}).

The models presented in this paper can be realized by fabricating an
$N$ terminal nanodevice. The heat conductance of suspended two-terminal
nanotubes was measured by detecting changes in the
electric resistance of the attached heater and detector pads
\cite{RectifE}. We suggest employing this method in a three-terminal setup,
 e.g. by connecting a T-shaped nanotube to three conducting surfaces.
Macromolecules with spatially localized sidegroups
also offer a playground for realizing these models.
\cite{Schwarzer1,Schwarzer2}.

Nonlinear electrical devices (rectifiers, switches and transistors)
have shaped technology in the last 60 years. Nonlinear
nanomechanical devices promise to revolutionize the technology of
the future as well, whereas phonons, instead of electrons become the
carriers of information and the computation element. This paper,
presenting a simple study for the control of  heat flow in nanosystems, is a
first step in addressing this challenge. Introducing quantum effects 
will further offer new pathways for energy control at the nanoscale \cite{Schwab1,Mingo}.

%




\begin{thebibliography}{32}


\bibitem{Uzer} 
T. Uzer, W. H. Miller, Phys. Rep. {\bf 199}, 73 (1991).

\bibitem{RRKM}
R. A. Marcus,
J. Chem. Phys. {\bf 20}, 359 (1952).

\bibitem{Wolynes}
M. Gruebele, P. G. Wolynes, Acc. Chem. Res. {\bf 37}, 261 (2004).

\bibitem{FPU}
E. Fermi, J. Pasta, S. Ulam, Los Alamos Document No. LA-1940 (1955).

\bibitem{FPURev}
D. K. Campbell, P. Rosenau, G. M. Zaslavsky,
Chaos {\bf 15}, 015101 (2005).


\bibitem{Kim}
P.  Kim , L. Shi, A.  Majumdar, P. L. McEuen, Phys. Rev. Lett. {\bf
87}, 215502 (2001).

\bibitem{Fujii}
M. Fujii,  {\it et al.},
Phys. Rev. Lett.  {\bf 95}, 065502  (2005).

\bibitem{Yu}
C. H. Yu, L. Shi, Z. Yao, D. Y. Li, A. Majumdar, Nano Lett. {\bf 5},
1842 (2005).

\bibitem{Braun}
Z. B.  Ge, D. G. Cahill, P. V. Braun, Phys. Rev. Lett. {\bf 96},
186101  (2006).

\bibitem{Segalman}
R. Y.  Wang, R. A. Segalman, A. Majumdar,
 App. Phys. Lett. {\bf 89},  173113 (2006).

\bibitem{Dlott}
Z. Wang, {\it et al.}, Science {\bf 317}, 787 (2007).

\bibitem{films}
C. Chiritescu, {\it et al.}, Science {\bf 315}, 351 (2007).

\bibitem{RectifE}
C. W. Chang, D. Okawa, A. Majumdar, A. Zettl, Science {\bf 314},
1121 (2006).


\bibitem{HeatSegal}
D. Segal, A. Nitzan,
J. Chem. Phys. {\bf 117}, 3915 (2002).

\bibitem{HeatDiVentra}
R. D'Agosta, N. Sai, M. Di Ventra,
Nano Lett. {\bf 6}, 2935 (2006).

\bibitem{Galperin}
M. Galperin, M. A. Ratner, A. Nitzan,
J. Phys.- cond. mat. {\bf 19}, 103201  (2007).


\bibitem{Saira}
O. P. Saira,  {\it et al.},
Phys. Rev. Lett. {\bf 99}, 027203 (2007).

\bibitem{Pump}
D. Segal, A. Nitzan,
Phys. Rev. E.  {\bf 73}, 026109 (2006).

\bibitem{Cahill} D. G. Cahill, K. Goodson, A. Majumdar,
J. Heat Transfer {\bf 124}, 223 (2002).

\bibitem{Blencowe}
M. Blencowe,
Phys. Rep. {\bf 395}, 159 (2004).

\bibitem{Majumdar}
D. Li, S. T. Huxtable, A. R. Abramson, A. Majumdar,
J. Heat Transfer,  {\bf 127} 108, (2005).

\bibitem{CasatiChaos}
G.  Casati, Chaos {\bf 15}, 015120  (2005).

\bibitem{Bonetto}
F. Bonetto, J. L. Lebowitz, and L. Rey-Bellet, {\it Fourier's law: A
challenge to theorists}, Mathematical Physics, A. Fokas, A.
Grigoryan, T. Kibble, and B. Zegarlinski, eds. (Imperial College
Press, London, 2000).

\bibitem{Lepri}
S. Lepri, R. Livi, A. Politi,
Phys. Rep. {\bf 377}, 1 (2003).

\bibitem{Gemmer}
M. Michel, G. Mahler, J. Gemmer,
Phys. Rev. Lett. {\bf 95}, 180602  (2005).

\bibitem{Pereira}
E. Pereira, R. Falcao,
Phys. Rev. Lett. {\bf 96}, 100601 (2006); F. Barros, H. C. F. Lemos,
E. Pereira, Phys, Rev. E {\bf 74}, 052102 (2006).


\bibitem{Casati1}
M. Terraneo, M. Peyrard,  G. Casati,
Phys. Rev. Lett. {\bf 88}, 094302 (2002).

\bibitem{Casati2}
B. Li, L. Wang, G. Casati,
Phys. Rev. Lett. {\bf 93}, 184301 (2004).


\bibitem{Rectif}
D. Segal, A. Nitzan, Phys. Rev. Lett. {\bf 94}, 034301 (2005);
J. Chem. Phys. {\bf 122}, 194704 (2005).


\bibitem{Rego} L. G. C. Rego,  G. Kirczenow,
Phys. Rev. Lett. {\bf{81}}, 232 (1998).



\bibitem{heatcond}
D. Segal, A. Nitzan,  P. Hanggi, J. Chem. Phys. {\bf 119}, 6840
(2003).





\bibitem{Casatixxx}
 B. W.  Li, L. Wang, G. Casati,
 App. Phys. Lett. {\bf 88}, 143501 (2006).

\bibitem{Tshape}
M. Menon, D. Srivastava,
Phys. Rev. Lett. {\bf 79}, 4453 (1997).

\bibitem{Trinkunas}
G. Trinkunas, A. Holzwarth,
J. Phys. Chem. B {\bf 101}, 7271 (1997).

\bibitem{Schwarzer2}
R. von Benten, O. Link, B. Abel, D. Schwarzer,
J. Phys. Chem. A {\bf 108}, 363 (2004).


\bibitem{Dhar}
Harmonic chains with self consistent reservoirs can also manifest normal heat
conductivity, see e.g.
F. Bonetto, J. L. Lebowitz, J. Lukkarinen,
J. Stat. Phys. {\bf 116}, 783 (2004);
A. Dhar, D. Roy,
J. Stat. Phys. {\bf 125}, 805 (2006).


\bibitem{NDR}
D. Segal, Phys. Rev. B {\bf 73}, 205415 (2006).

\bibitem{Lin}
S. H. Lin, J. Chem. Phys. {\bf 61}, 3810 (1974).

\bibitem{Peierls} R. E. Peierls, {\it Quantum Theory of Solids} (Oxford University
Press, London, 1955).

\bibitem{Mahan} G. D. Mahan, {\it Many-particle physics} (Plenum press, New York, 2000).

\bibitem{Carmeli}
B. Carmeli, A. Nitzan, Chem. Phys. Lett. {\bf102}, 517 (1983).

\bibitem{Schwarzer1}
D. Schwarzer, P. Kutne, C. Schroder, J. Troe,
J. Chem. Phys. {\bf 121}, 1754 (2004).

\bibitem{Schwab1}
K. C. Schwab, M. L. Roukes,
Phys. Today {\bf 58}, 36 (2005).

\bibitem{Mingo}
N. Mingo, Phys. Rev. B {\bf 74}, 125402 (2006).

\end{thebibliography}
\end{document}